\documentclass[usenatbib]{mnras}

\usepackage{newtxtext,newtxmath}

\usepackage[T1]{fontenc}
\usepackage{ae,aecompl}

\usepackage[utf8]{inputenc}
\usepackage[T1]{fontenc}

\usepackage{graphicx}	
\usepackage{amsmath}	
\usepackage{systeme}    

\title[The magnetic $\delta$\,Scuti star HD\,41641]{\Huge The complex fossil magnetic field of the $\delta$\,Scuti star HD\,41641}

\author[K. Thomson-Paressant et al.]{
K. Thomson-Paressant,$^{1}$
C. Neiner,$^{1}$\thanks{E-mail: coralie.neiner@obspm.fr}
K. Zwintz$^{2}$
and A. Escorza$^{3,4}$
\\
$^{1}$LESIA, Paris Observatory, PSL University, CNRS, Sorbonne University, Université de Paris, 5 place Jules Janssen, 92195\\ Meudon, France\\
$^{2}$Institut f\"ur Astro- und Teilchenphysik, Universit\"at Innsbruck, Technikerstraße 25, 6020 Innsbruck, Austria\\
$^{3}$Institute of Astronomy, KU Leuven, Celestijnenlaan 200D, 3001, Leuven, Belgium\\
$^{4}$Institut d'Astronomie et d'Astrophysique, Universit\'{e} Libre de Bruxelles (ULB), CP 226, B-1050 Bruxelles, Belgium}

\date{Accepted XXX. Received YYY; in original form ZZZ}

\pubyear{2020}

\begin{document}
\label{firstpage}
\pagerange{\pageref{firstpage}--\pageref{lastpage}}
\maketitle

\begin{abstract}
Only three magnetic $\delta$\,Scuti stars are known as of today. HD\,41641 is a $\delta$\,Scuti star showing chemical peculiarities and rotational modulation of its light-curve, making it a good magnetic candidate. We acquired spectropolarimetric observations of this star with Narval at TBL to search for the presence of a magnetic field and characterize it. We indeed clearly detect a magnetic field in HD\,41641, making it the fourth known magnetic $\delta$\,Scuti star. Our analysis shows that the field is of fossil origin, like magnetic OBA stars, but with a complex field structure rather than the much more usual dipolar structure.
\end{abstract}




\section{Introduction}

$\delta$\,Scuti ($\delta$\,Sct) variables are stars exhibiting variations in their luminosity from low radial order pressure modes and mixed modes, which are themselves driven by a combination of the $\kappa$ mechanism operating within the He II ionisation zone \citep[][]{pamyatnykh2000} coupled with turbulent pressure in the hydrogen ionisation layer \citep[][]{Antoci2014,xiong2016}. These pulsations typically last from 15 min to up to 8 h. $\delta$\,Sct stars span the spectral range A through F, with masses generally of the order of 1.5-4.0 M$_\odot$ \citep[][]{aerts2010} and effective temperatures between $\sim$6700 and 8000 K.

To date, three $\delta$\,Sct variables have been found to exhibit magnetic fields: HD\,188774 \citep[][]{neiner2015}, HD\,67523 \citep[][]{neiner2017}, and $\beta$\,Cas \citep{zwintz2019,zwintz2020}. HD\,21190 has also been claimed to be a  magnetic $\delta$\,Sct star \citep{Kurtz2008, HubrigScholler2016}, however this result is in contention with those from \cite{Bagnulo2012} and thus is still up for debate. Finally, HD\,35929 was claimed as a possible magnetic $\delta$\,Sct candidate  \citep{Alecian2013} but the spectropolarimetric results were unclear and have not been confirmed.

Pulsating stars, such as $\delta$\,Sct and $\beta$\,Cep variables, are important stars in the field of asteroseismology, as their oscillations allow researchers to determine their internal structure and expand our understanding of the structure and evolution of stars. In the case of $\delta$\,Sct-type stars, the oscillation modes are low-order radial and non-radial pressure modes. The presence of magnetic fields within the star, as is the case with certain $\delta$\,Sct stars, allow us to put extra constraints on the modelling of this internal structure \citep[e.g.][]{buysschaert2018, prat2019, prat2020}. For example, even a relatively weak field at the surface is enough to inhibit mixing inside the star and thus the overshoot parameter can be put at 0 in the seismic model \citep{briquet2012}.

HD\,41641 ($\alpha_{2000}=6^\text{h}6^\text{m}40.578^\text{s}, \delta_{2000}=+6^\circ{}43'49.886"$) is a prime example of a $\delta$\,Sct star, with a mass of $2.3\pm0.5 \; M_\odot$ and an effective temperature of $7200\pm80 \; K$ \citep[][hereafter E16]{escorza2016}. It is a bright ($V=7.86$\,mag), A5III-type star that is well placed within the $\delta$\,Sct instability strip of the Hertzprung-Russell diagram. E16 detected a number of frequencies associated with oscillation modes present within the star, with periods ranging between 10 and 20 $\text{d}^{-1}$, and postulated the existence of a magnetic field because of the presence of chemical peculiarities. Table~\ref{tab:info} displays the key information about HD\,41641 as determined by E16. During their investigation, E16 determined two possible frequencies that most likely relate to the rotation period of the star and its harmonic, labelled F7 and F24, with preference for the rotation itself  given to the latter.

In this paper, we search for the presence of a magnetic field in HD\,41641 using spectropolarimetry, to confirm HD\,41641 as the fourth known magnetic $\delta$\,Sct star. 

The observations of HD\,41641 are summarised in Sect.~\ref{section2}, the measurements of a magnetic field are presented and modelled in Sect.~\ref{section3}, and the results are discussed in Sect.~\ref{section4}.

\section{Spectropolarimetric Observations and Methods}
\label{section2}

\subsection{Observations}

To check for the existence of a magnetic field around HD\,41641, it was observed using the Narval echelle spectropolarimeter \citep[][]{narval2003}, which is operating on the Bernard Lyot Telescope (TBL) at the Observatoire Midi-Pyr\'en\'ees, France.

HD\,41641 was observed over 17 nights between 21 October 2015 and 2 December 2016. The list of the nightly average observations is visible in Table~\ref{tab:observations}, with all the individual observations presented in Table~\ref{tab:observations2} in the Appendix.

\begin{table}
	\centering
	\caption{Stellar parameters as determined by \citet{escorza2016}.}
	\label{tab:info}
	\begin{tabular}{rl|rl}
		\hline
		$V$ & 7.86\,mag & $\log g$ & $3.5\pm 0.3$ dex \\
		$M_*$ & $2.3^{+0.7}_{-0.5} \; M_\odot$ & $v \sin i$ & $30\pm 2 \; \text{km.s}^{-1}$ \\
		$T_{\rm eff}$ & $7200\pm80$ K & RV & $29.3\pm0.3 \; \text{km.s}^{-1}$ \\
		$R_*$ & $4.5^{+2.7}_{-1.7} \; R_\odot$ & $\xi$ & $1.1\pm0.3 \; \text{km.s}^{-1}$\\
		\hline
	\end{tabular}
\end{table}

We used the circular polarisation mode to measure the Stokes V spectrum along with the intensity spectrum, also known as Stokes I. Each Stokes V profile consists of the combination of 4 sub-exposures, which are obtained with the polarimeter's half-wave Fresnel rhombs set at various angles. The sub-exposures are also destructively combined to produce a null polarisation spectrum, labelled N, to check for pollution by factors such as variable observing conditions, instrumental effects, or non-magnetic stellar effects such as pulsations. In addition, successive Stokes V sequences were acquired to increase the overall signal-to-noise ratio (SNR) of a potential magnetic measurement.

The Least-Squares Deconvolution (LSD) method \citep[][]{donati1997} was used to average the spectral line profiles and the Stokes V profiles of each measurement. This technique works by combining all available lines in the spectrum, with a weight that depends on the Land\'e factor $g_L$ of the line, i.e. its sensitivity to the presence of a magnetic field, and the depth of the line. This improves the SNR of the combined line compared to a single line. LSD assumes the weak field hypothesis and that all lines used in the combination have the same shape.

To compute the necessary line mask, lines were selected from the VALD3 database \citep[][]{piskunov1995,kupka1999,ryabchikova2015}, which provides atomic data for stellar atmospheres. For HD\,41641, the initial mask was computed with an effective temperature $T_{\rm eff}=7250$ K and a gravity $\log g=3.5$ dex, selecting only lines with a depth of at least 1\% of the continuum level. From this mask, we then removed all hydrogen lines and any lines blended with either telluric lines or hydrogen lines, as well as lines present in the original VALD line list that did not appear in the observed spectrum. Through this processing, the total number of lines to be used by the LSD was reduced from an initial 13,979 to 6,745. For the final mask, the values for the mean Landé factor and mean wavelength were $g_L = 1.6807$ and $\lambda = 531.7517 \; \text{nm}$ respectively.
\newline

\begin{table}
\centering
\begin{tabular}{lcccc}
\hline
Date      & Mid-HJD   & T$_{\text{exp}}$ & Phase   & SNR  \\
          & -2450000  & s         &         &      \\
\hline
21 Oct 15 & 7317.6957 & 8x4x27    & 0.248 & 860  \\
09 Nov 15 & 7336.7578 & 3x4x27    & 0.034 & 571  \\
11 Nov 15 & 7338.7326 & 5x4x27    & 0.737 & 1325 \\
12 Nov 15 & 7339.7545 & 8x4x27    & 0.101 & 1610 \\
16 Nov 15 & 7343.6598 & 8x4x27    & 0.491 & 1666 \\
15 Mar 16 & 7463.3385 & 8x4x27    & 0.097 & 962  \\
20 Mar 16 & 7468.3606 & 7x4x27    & 0.884 & 1846 \\
08 Oct 16 & 7670.6344 & 10x4x27   & 0.894 & 2129 \\
27 Oct 16 & 7689.5768 & 10x4x27   & 0.637 & 683  \\
28 Oct 16 & 7690.7037 & 10x4x27   & 0.038 & 1763 \\
29 Oct 16 & 7691.6058 & 10x4x27   & 0.360 & 1775 \\
30 Oct 16 & 7692.7331 & 10x4x27   & 0.761 & 1808 \\
31 Oct 16 & 7693.6274 & 10x4x27   & 0.079 & 2231 \\
02 Nov 16 & 7695.6581 & 10x4x27   & 0.802 & 1843 \\
25 Nov 16 & 7718.6306 & 10x4x27   & 0.980 & 1841 \\
01 Dec 16 & 7724.4852 & 10x4x27   & 0.065 & 1161 \\
02 Dec 16 & 7725.6170 & 10x4x27   & 0.468 & 2237 \\
\hline
\end{tabular}
\caption{Summarised journal of observations of HD\,41641, with averaged nightly profiles and indications of mean Julian Date at midpoint of each observation, exposure time, phase (assuming F7 as the rotation frequency) and signal-to-noise ratio. The chosen value for HJD$_0$ is 2457317.0.}
\label{tab:observations}
\end{table}

To improve the SNR further and reduce the impact of potential remnant signals in N, we performed a weighted average on the observations for each night, in order to generate a mean profile per night which would be used for the remainder of the analysis process. On average, the LSD improved the SNR by a factor of almost 17x between the spectrum and the LSD profile at 500 nm. By averaging several observations, the SNR between a single measurement and the nightly average can be improved by a further 3x, giving a final improvement of 51x.

\begin{figure}
	\includegraphics[width=\columnwidth]{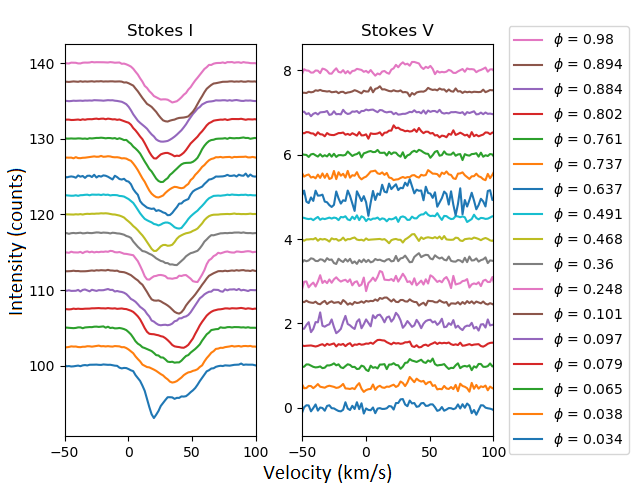}
    \caption{Stokes I (left) and Stokes V (right) LSD profiles of HD\,41641 for the averaged 17 nights of observation, artificially shifted vertically to improve readability and ordered from bottom to top with respect to their phase using the rotation frequency F7.}
    \label{fig:stokes}
\end{figure}

Through our analysis, it was determined that the observations on 27 Oct 2016 were poor, having a residual signal in N that remained even after night-by-night averaging. This is visible in Fig.~\ref{fig:stokes}, with the night in question being represented by the dark blue line at phase 0.637. Numerous attempts were made to improve the profile for this night, such as assessing each observation individually and looking for defects, but no clear improvement was observed, and thus it was decided to not consider the 27 Oct 2016 for the remainder of the work.

As mentioned in the introduction, E16 discovered two potential frequencies most likely related to the rotation period of HD\,41641 labelled F7 and F24, corresponding to $2.80898876 \; \text{d}$ and $5.63189907 \; \text{d}$ respectively. The authors concluded that the rotation period was best represented by F24, but for the sake of completeness and accuracy, both were tested here. As will be discussed further later in the paper, F7 was determined to be the correct rotation frequency, and thus all figures and tables displayed here are with respect to this frequency.

\subsection{Testing chemical signatures}

The LSD Stokes I profiles visible in the left panel of Fig.~\ref{fig:stokes} displayed bump features present within the line profile that evolve on a night-by-night basis. This could be the result of either brightness or chemical spots on the surface of the star, or the $\delta$\,Scuti pulsations. In the case of a spot, this can be tested by comparing the bumps by particular chemical elements in order to determine their distribution on the surface.

With this in mind, we created a submask for each element present within the LSD mask and which had $\gtrsim 30$ lines to maintain a minimum SNR threshold (with a few exceptions such as Nitrogen). We ended up with 24 submasks, each with a number of respective lines as presented in Table \ref{tab:elements}. Since the lines in the original mask were extracted from VALD3, the line identification used for the submask is the one from VALD3, which may not always be correct. Due to the limited number of lines in the case of certain elements, we chose to group lines resulting from both neutral and ionised states of each element together.

By applying the LSD method with each of the chemical submasks, we can generate a set of profiles that correspond to the lines from each chemical element, which can then be averaged night-by-night, as was done for the full spectrum. These can then be analysed in two ways: either by comparing for a single element all the averaged nightly profiles, or by comparing on a single night the averaged spectra for all the elements. 

We elected to begin by the latter, observing all the elements for a single night, beginning with the 2 Dec 2016 due to its high SNR, and eventually comparing individual nights. We observed that most elements displayed a clear "dip" on the blue wing of the profile and additional smaller features, while others had less clear features with the strongest sometimes rather on the red wing. This initially seemed to define two families of elements, suggesting the possible existence of two distinct chemical regions on the surface. To confirm this, the elements were grouped into their two respective families and plotted by family. It was determined that those presenting a main feature on the red wing simply had too low of an SNR, presenting other noisy features not related to anything physical. Indeed, by referring back to the number of lines that these elements presented, they all had $<100$ lines each. As such, it was concluded that only a single pattern exists, with a main spot on the blue wing on 2 Dec 2016, and visible in all chemical elements for which sufficient SNR can be reached. Checking the other nights, we saw that this pattern travels along the line profiles with time. Since it appears for all chemical elements, it must be either a brightness pattern or pulsations, rather than specific chemical inhomogeneities.

\begin{table}
\centering
\caption{Distribution of spectral lines for elements detected within the spectra of HD\,41641.}
\label{tab:elements}
\begin{tabular}{rllc}
\hline
Z & \multicolumn{2}{c}{Element} & Nb of lines \\
\hline
6         & Carbon       & C      & 119         \\
7         & Nitrogen     & N      & 11          \\
8         & Oxygen       & O      & 21          \\
11        & Sodium       & Na     & 23          \\
12        & Magnesium    & Mg     & 64          \\
14        & Silicon      & Si     & 133         \\
16        & Sulfur       & S      & 61          \\
20        & Calcium      & Ca     & 125         \\
22        & Titanium     & Ti     & 311         \\
23        & Vanadium     & V      & 137         \\
24        & Chromium     & Cr     & 427         \\
25        & Manganese    & Mn     & 175         \\
26        & Iron         & Fe     & 1763        \\
27        & Cobalt       & Co     & 113         \\
28        & Nickel       & Ni     & 189         \\
39        & Yttrium      & Y      & 30          \\
40        & Zirconium    & Zr     & 55          \\
57        & Lanthanum    & La     & 41          \\
58        & Cerium       & Ce     & 344         \\
59        & Praseodymium & Pr     & 40          \\
60        & Neodymium    & Nd     & 164         \\
62        & Samarium     & Sm     & 111         \\
64        & Gadolinium   & Gd     & 64          \\
66        & Dysprosium   & Dy     & 36          \\
\hline
\end{tabular}
\end{table}

In E16, during their analysis of the chemical abundances of HD\,41641, the authors suggested that these spot-like features might be due to local concentrations of particular elements. However, from our results, we must rather conclude that the spots are not related to local concentrations of chemical elements.

\subsection{Testing the presence of a companion}

A theoretical companion to HD\,41641 had been proposed by both E16 and \citet{poretti2005}, but both achieved different conclusions after further study. Poretti suggested that HD\,41641 could be a double-lined spectroscopic binary, whereas E16 determined that other lines within the frequency space were merely due to chemical spots.

In the spectra presented here, we note a small velocity shift (visible in Fig.~\ref{fig:compplot}) between  observations taken 6 months apart. This does not seem to be related to an  instrumental effect and could therefore be a sign of the presence of a companion. In addition, through the analysis of the intensity profiles by chemical element, a feature was discovered with a slight offset in velocity to the primary source, between -100 and -50 km/s, which could also be a signature from a companion. We subsequently checked for the existence of a companion to our target star. 

We theorised that, should such a companion exist, and considering that it does not appear in the full LSD profiles (i.e. with all chemical elements included), it would be much cooler than our target. Thus, we performed a similar process as had been completed up to now with our target, but instead using a line mask extracted from the VALD3 database information for significantly cooler stars, setting $T_{\rm eff}=3500 \;\text{K}$ and maintaining an identical value for the surface gravity. If a cool companion was present, this process should accentuate the lines from the companion and reveal its existence. Having completed this analysis, it was quickly apparent that no clear signs of a companion exist. In particular the feature between -100 and -50 km/s does not appear in the cool star LSD profiles and was likely just noise.

To summarise, in the data presented in this study, we find no compelling evidence of a cool companion. 

\section{Magnetic field}
\label{section3}

\subsection{Magnetic field measurements}

The LSD Stokes V profiles visible in the right panel of Fig.~\ref{fig:stokes} display a clear signature. Applying a False Alarm Probability (FAP) algorithm to the averaged nightly profiles yielded that 10 out of the 17 nights revealed definite signal detection in V, with an additional one showing a marginal detection. A definite detection is defined by a value smaller than $\text{FAP} \lesssim 10^{-5}$, and marginal by $10^{-5} \lesssim \text{FAP} \lesssim 10^{-3}$  \citep{donati1997}. There were no detections in N for any of the profiles. The number of definite detections in Stokes V, with none in N, clearly shows the existence of a magnetic signal in HD\,41641. Although the FAP shows no detection in N, we decided nonetheless to continue to ignore the 27 Oct 16 profile in the remainder of this work, as visibly its N profile does not look flat (see the profile for phase 0.637 in Fig.~\ref{fig:stokes}).

From the remaining 16 averaged profiles, after having removed the 27 Oct 16 data, dynamical plots were generated, with a phase binning of 0.017. The bin size was selected in such a way as to ensure the greatest amount of clarity for the dynamical plots, and reduce empty space between profiles. As such, some profiles overlap slightly in phase-space.

\begin{figure}
	\resizebox{\hsize}{!}{\includegraphics[clip]{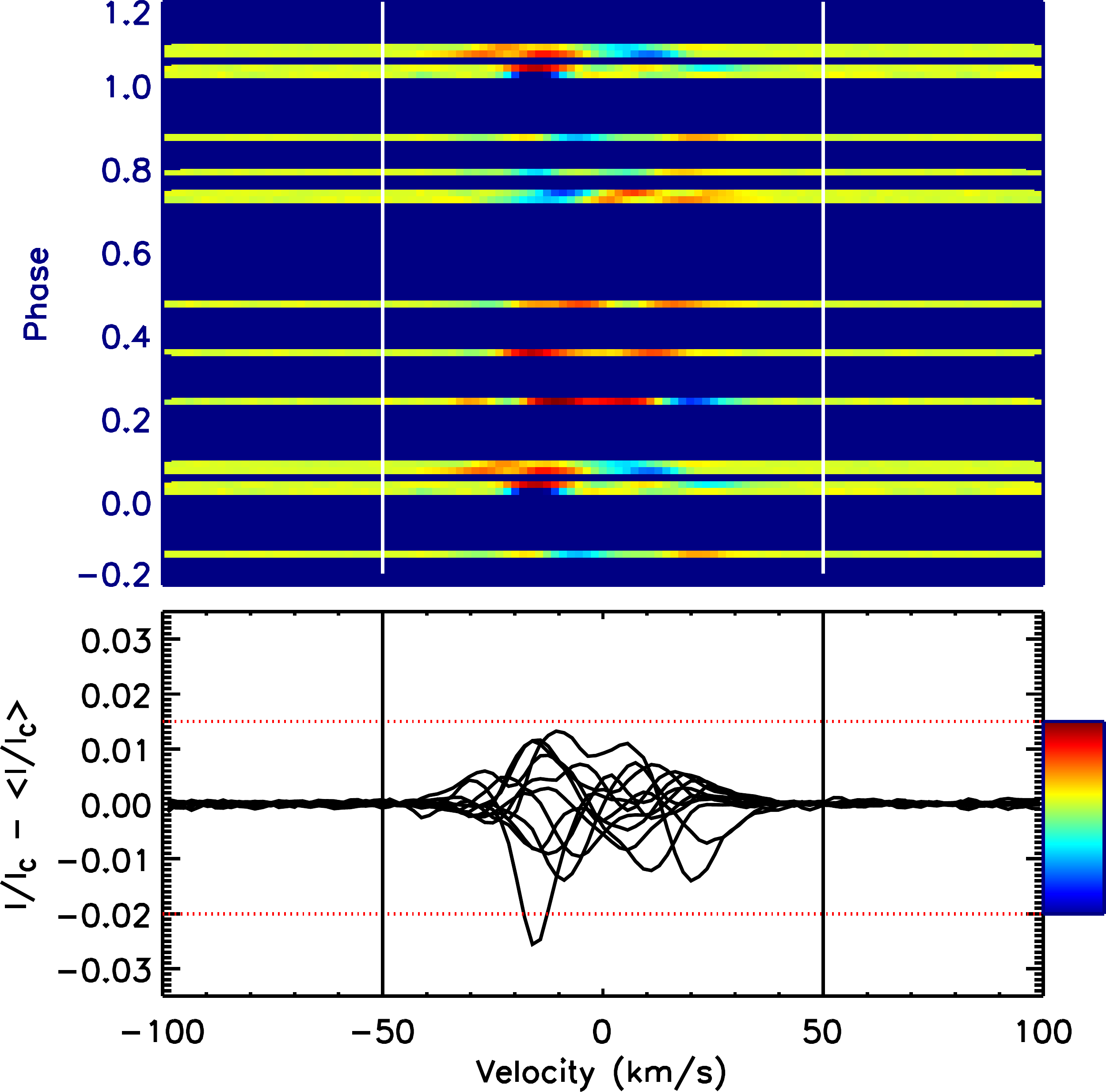}}
    \caption{Residuals of the LSD Stokes I profiles with respect to the averaged Stokes I profile shown over-plotted (bottom panel) and as a dynamical plot folded in phase with the rotation frequency F7 (top panel). Colourbar represents values of Stokes I that vary between -0.02 (blue) and 0.015 (red) of the average profile value.}
    \label{fig:stokesI_dynplot}
\end{figure}

\begin{figure}
	\resizebox{\hsize}{!}{\includegraphics[clip]{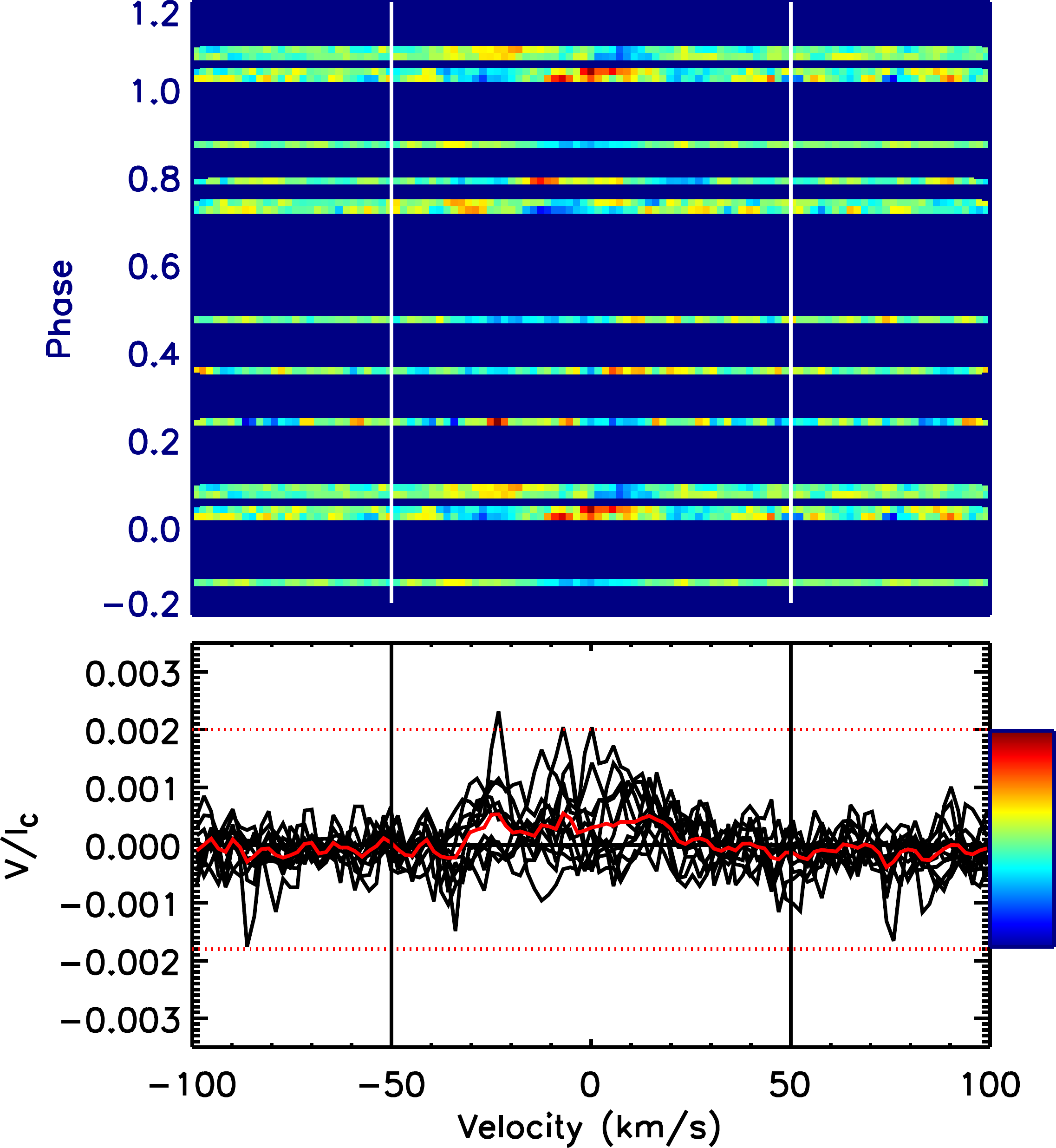}}
    \caption{LSD Stokes V profiles (bottom panel) and corresponding dynamical plot folded in phase with the rotation frequency F7 (top panel). Colourbar represents values of Stokes V that vary between -0.0018 (blue) and 0.002 (red). The red line corresponds to the average of all the profiles.}
    \label{fig:stokesV_dynplot}
\end{figure}

Fig.~\ref{fig:stokesI_dynplot} and \ref{fig:stokesV_dynplot} display the residual LSD Stokes I and the Stokes V profiles respectively in the bottom panel, as well as colourscales of these profiles folded with the frequency F7 in the top panel. 

For Stokes I (Fig.~\ref{fig:stokesI_dynplot}), although the phase coverage is not ideal, the evolution of the bumps along the profile is nonetheless visible. This pattern is visible when the profiles are folded with F7, i.e. with the rotation period or an harmonic thereof. Nothing clear appears when we fold the Stokes I profiles with any of the main pulsation frequencies listed by E16. We thus conclude that the observed bumps are due to a brightness pattern rather than to the $\delta$\,Scuti pulsations.

This brightness pattern could be directly related to the presence of an oblique dipole field, as observed in hot stars, for example the main spot could be a magnetic pole or there could be an eclipse by a magnetosphere. However, looking at the Stokes V colourscale (Fig.~\ref{fig:stokesV_dynplot}), there is no clear dipolar pattern visible.

\subsection{Longitudinal magnetic field}
\label{subsec:longitudinal_field}

\begin{figure}
	\includegraphics[width=\columnwidth]{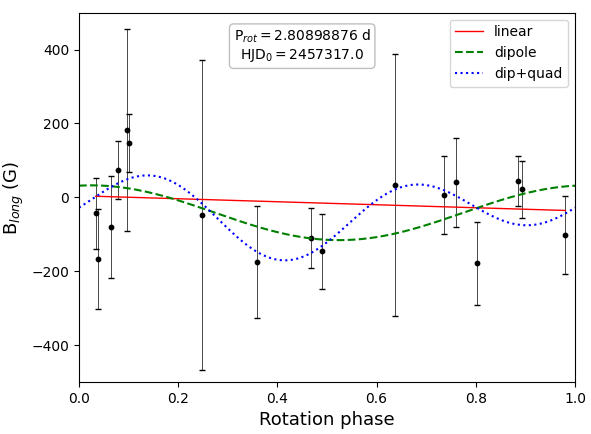}
    \caption{$B_l$ values with respect to phase, setting F7 as the rotation frequency. Full red, dashed green and dotted blue line fits represent respectively the linear, sinusoidal (dipolar field) and double-wave (dipolar+quadrupolar field)  models of the $B_l$ values.}
    \label{fig:Bl_plot}
\end{figure}

The next stage of the process involves calculating the values for the longitudinal magnetic field $B_l$ from the LSD Stokes V and I profiles. It is important to define an appropriate region about the line profile, as selecting one that is too large or too small can cause erroneous results. With this in mind, we determined a velocity range of $\pm 50\; \text{km.s$^{-1}$}$ around the line centroid. The results are displayed in Table~\ref{tab:Bvalues}, along with similar measurements in $N$, and their respective error values. Subsequently these $B_l$ values were plotted with respect to rotation (F7) phase as seen in Fig.~\ref{fig:Bl_plot}, along with a 1-term and a 2-term sinusoidal fit, in order to visualise the variation of the magnetic field with respect to the rotation of the star. We find that the $B_l$ values do not seem to follow a sinusoidal pattern, i.e. that the magnetic field does not seem to be dipolar. 

To get a sense of the magnetic field strength, we nevertheless use the dipolar fit to $B_l$  values and deduce:


\[\vspace{10pt} \left\{\begin{aligned}
B^{+}_{l} &\sim 59.6 \pm 0.6 \; \text{G} \\ 
B^{-}_{l} &\sim -170.3 \pm 0.6 \; \text{G}
\end{aligned}\right.
\]

We can then derive the polar field strength $B_{\rm pol}$ and the magnetic field obliquity $\beta$ (assuming a dipole field).

Since we determined that F7 is the rotation frequency rather than F24, we first determined a new value for $i$, which differ from the one found by E16. This value corresponds to $i=22 \pm 11^\circ$. Using the ratio $r = B^{+}_{l} / B^{-}_{l}  \sim -0.35$ and following the work completed by \cite{preston1967} and \cite{landstreet1970}, we then find  $\beta= 79 \pm 11^\circ$. Finally, a value for $B_{\rm pol}$ can be determined following \cite{schwarzschild1950} and using a limb-darkening coefficient of 0.4:

\begin{equation}
    B_{\rm pol} = \frac{B^{\pm}_{l}}{0.296\times \text{cos}(\beta \pm i)}
\end{equation}

Using the aforementioned parameters for $i$, $\beta$ and $B_l$, we calculated that $B_{\rm pol}$ is 1055 G. Considering error bars it could be as low as 366 G and no maximum value can be derived.

In addition, we modeled the LSD Stokes V profiles with these parameters. We obtained a poor fit to the observations, in particular an incompatible shape of the Stokes V profiles, which confirms that a simple dipole does not match the observations. We can then compare the reduced $\chi^2$ values for a linear ($\chi^2_{\rm red}=1.098$), dipolar ($\chi^2_{\rm red}=0.772$) as well as dipolar+quadrupolar ($\chi^2_{\rm red}=0.887$) fit. These results suggest that the dipolar and dipolar+quadrupolar fits are slightly better fits to the data than the linear fit. This suggests that some rotational modulation is indeed present in the data. However, the error bars for some points are particularly large and could be represented by any fit, while several other points are wholly excluded from the sinusoidal fits. We therefore conclude that rotational modulation is present in the data but that a simple configuration (dipolar or dipolar+quadrupolar) does not represent the data well.

\begin{table}
\centering
\begin{tabular}{cccc}
\hline
Date & FAP & $B_l \pm \sigma B$ (G) & $N_l \pm \sigma N$ (G)\\
\hline
21 Oct 15 & $2.3\times 10^{-1}$ (N) & $-47\pm 420$  & $-30\pm 420$ \\
9 Nov 15  & $1.4\times 10^{-2}$ (N) & $-43\pm 97$   & $29\pm 97$   \\
11 Nov 15 & $1.2\times 10^{-2}$ (N) & $6\pm 105$    & $-16\pm 105$ \\
12 Nov 15 & $2.7\times 10^{-6}$ (D) & $148\pm 79$   & $5\pm 79$    \\
16 Nov 15 & $2.8\times 10^{-3}$ (N) & $-145\pm 102$ & $-28\pm 102$ \\
15 Mar 16 & $3.6\times 10^{-2}$ (N) & $182\pm 273$  & $17\pm 272$  \\
20 Mar 16 & $6.5\times 10^{-2}$ (N) & $45\pm 68$    & $21\pm 68$   \\
8 Oct 16  & $4.6\times 10^{-7}$ (D) & $22\pm 78$    & $-45\pm 78$  \\
27 Oct 16 & $1.5\times 10^{-6}$ (D) & $33\pm 355$   & $-31\pm 355$ \\
28 Oct 16 & $1.8\times 10^{-15}$ (D) & $-116\pm 134$ & $102\pm 135$ \\
29 Oct 16 & $6.0\times 10^{-4}$ (M) & $-175\pm 151$ & $-28\pm 151$ \\
30 Oct 16 & $1.1\times 10^{-5}$ (D) & $41\pm 120$   & $-72\pm 120$ \\
31 Oct 16 & $8.7\times 10^{-12}$ (D) & $74\pm 78$    & $30\pm 78$   \\
2 Nov 16  & $1.3\times 10^{-10}$ (D) & $-178\pm 112$ & $-66\pm 112$ \\
25 Nov 16 &             $0.0$ (D)   & $-102\pm 105$ & $6\pm 105$   \\
1 Dec 16  & $2.8\times 10^{-8}$ (D) & $-81\pm 138$  & $36\pm 138$  \\
2 Dec 16  & $7.4\times 10^{-9}$ (D) & $-109\pm 81$  & $-16\pm 81$  \\
\hline
\end{tabular}
\caption{Magnetic detection status from False Alarm Probability (FAP) results, along with longitudinal magnetic field measurements ($B_l$) from the LSD Stokes V profiles and similar measurements ($N_l$) from the LSD Null profiles, including their respective errors calculated outside the line profile region. Also included are False Alarm Probability (FAP) results, with magnetic detection status represented by Definite (D), Marginal (M) and Non-detection (N).}
\label{tab:Bvalues}
\end{table}

\subsection{Choice of rotation frequency}

As we mentioned earlier, there was some disagreement between our findings and those of E16 with regards to which frequency corresponded to the rotation frequency of HD\,41641. E16 suggested that the frequency labelled F24, with a value of $f=0.17756 \; \text{d}^{-1}$, was the correct one whereas our results suggest that F7, at $f=0.3560 \; \text{d}^{-1}$, coincides more accurately with the rotation. Both frequencies are intrinsically linked, with F7 being twice that of F24. 

\begin{figure}
	\includegraphics[width=\columnwidth]{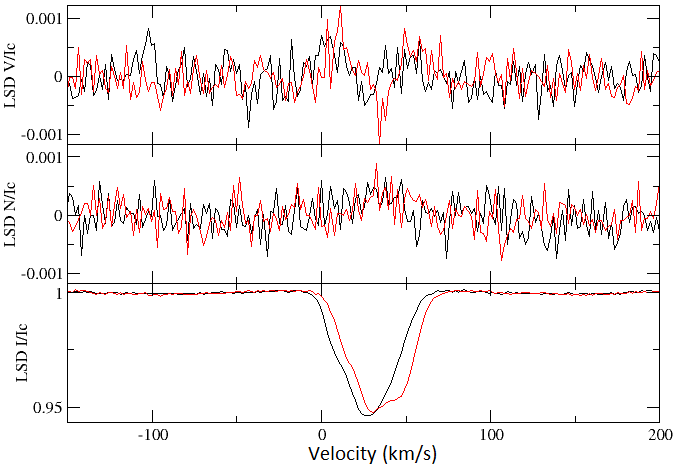}
    \caption{Stokes V (top), N (middle), and Stokes I (bottom) LSD profiles of HD\,41641 for the nights of the 20 Mar 2016 (black) and 8 Oct 2016 (red).}
    \label{fig:compplot}
\end{figure}

We determined F7 to be the correct rotation frequency due to the following factors:
\begin{itemize}
    \item By plotting the $B_l$ values with respect to F7, as seen in Fig.~\ref{fig:Bl_plot}, we see that observations obtained at different epochs but similar phases have similar longitudinal field values. Although the signal is not sinusoidal (i.e. dipolar), it appears periodic with F7. Using F24 instead shows no such periodicity.
    \item Similarly, this is also the case when considering the residual Stokes I dynamical plot, Fig.~\ref{fig:stokesI_dynplot}, where the evolution of the line profile with respect to phase draws distinct tracks with both the positive (blue) and negative (red) amplitudes when using F7. Again, this is not apparent when substituting F24.
    \item Finally, comparing Stokes V profiles of observations taken approximately 6 months apart (Fig.~\ref{fig:compplot}) on 20 Mar 2016 and 8 Oct 2016, and with very similar phase values of 0.88436 and 0.89384 respectively when using F7, it is immediately apparent that they are analogous apart from a slight shift in velocity visible in both Stokes I and V. These 2 similar profiles would have different phases when using F24.
\end{itemize}

\subsection{Modelling the LSD Stokes profiles}

To further check the configuration of the magnetic field, an attempt was made to model the LSD Stokes V profiles with a dipole field with free parameters. To this end, we utilised the method described in \citet{alecian2008}. In essence, this is an oblique rotator model, where we fit 4 or 5 parameters: the inclination angle $i$ between the rotation axis and observer's line of sight, the obliquity angle $\beta$ of the dipole field with respect to the rotation axis, the polar field strength $B_{\rm pol}$, a phase shift with respect to the chosen ephemeris $\Delta \psi$, and an optional off-centring distance $d$ in the off-centred dipole case ($d=0$ for a centred dipole and $d=1$ if the centre of the dipole is at the surface of the star). We provide certain values such as macro- and micro-turbulence, phase values assuming F7 is the rotation frequency, mean wavelength and Land\'{e} factor, while fitting Gaussians to the Stokes I spectra to generate values for the depth of the Stokes I profile, $v \sin i$, and the radial velocity.

We calculated a grid of Stokes V profiles by varying the 4 or 5 parameters within user-defined bounds, and using a $\chi^2$ minimisation to determine the best fit values. This can be repeated several times, with the parameters converging on more accurate values with each loop. Using these best fit parameters, we can then try to fit a model curve to all the LSD Stokes V profiles simultaneously, in an attempt to reveal a possible dipolar signal. 

It became quickly apparent that no such dipolar signal exists for HD\,41641, as the model fits extremely poorly, fitting extremely few of the features visible in the LSD profiles. To ensure that the fitting process had converged properly, it was repeated several times with various first guess parameters, while also rejecting the region of the main spot within the profile when fitting the Stokes I spectra prior to modelling.

The results of the modelling are listed as follows. We determined two combinations of inclination and obliquity angles, which effectively corresponds to $i=70\pm 5^\circ$ and $\beta = 30 \pm 5^\circ$, and $i=30\pm 5^\circ$ and $\beta = 70 \pm 5^\circ$, both with $\chi^2=1.875$. The model also suggests a phase shift $\Delta\psi = 0.0738\pm 0.005$ and, assuming it is a dipole, a maximum polar field strength of $B_{\rm pol} = 1880 \pm 10 \; \text{G}$. 

\begin{figure*}
	\includegraphics[width=12cm]{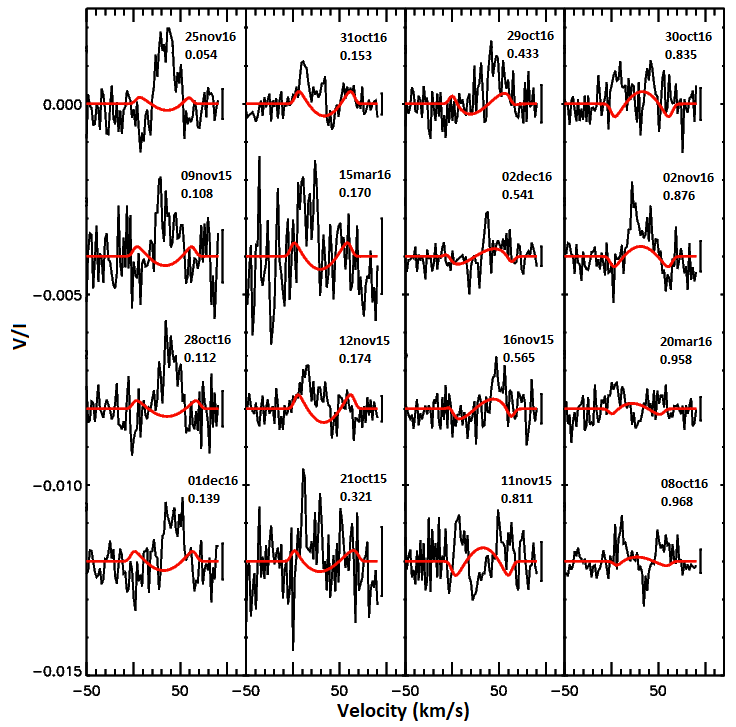}
    \caption{Results of dipolar modelling performed on the 16 best profiles. Profiles are ordered in phase, increasing vertically first then horizontally. Model fit is displayed as red overlay.}
    \label{fig:dipmod}
\end{figure*}

The best-fit model is visible in Fig.~\ref{fig:dipmod}. We stress that the dipolar fit to the data is very poor and the above values must therefore be considered with great care and are provided here only to give an idea of the amplitude of the magnetic field signature. This model again shows that a dipolar fit is not appropriate for this star.

Comparing the results determined here for $i$, $\beta$ and $B_{\rm pol}$ with those calculated in Sect.~\ref{subsec:longitudinal_field}, it is clear there is quite an offset, though this is expected as a result of the poor fitting dipolar modelling. We do however see a match between the calculated values of $\beta=79\pm 11^\circ$ and $i=22 \pm 11^\circ$ and the model with values of $\beta=70\pm 5^\circ$ and $i=30\pm 5^\circ$. Finally, the error bar of the calculated $B_{\rm pol}$ looks far more reasonable than the range of possible values determined by the dipolar modelling in  Sect.~\ref{subsec:longitudinal_field}, but considering the bad fit we obtain, this error bar is not realistic.

\section{Conclusions}
\label{section4}

We carried out a spectropolarimetric study of the $\delta$ Sct star HD\,41641, complementing the findings of E16. We utilised data taken with the Narval echelle spectropolarimeter from 21 October 2015 to 2 December 2016. The spectra were used to determine the existence and possible origin of the magnetic field of HD\,41641, as well as confirm some of the stellar parameters as determined by E16. 

First of all, we confirm that HD\,41641 is indeed magnetic, making it the 4th magnetic $\delta$\,Sct star discovered to date. Through our subsequent analysis, we determined that a different rotation frequency than the one suggested by E16 fit the data more accurately. Switching from a frequency of $0.17756 \; \text{d}^{-1}$ to $0.3560 \; \text{d}^{-1}$ revealed a periodic variation in the magnetic field measurements as well as in the LSD Stokes I profiles. As a result, some of the other stellar parameters calculated by E16, such as the inclination $i$, were no longer suitable and required recalculation. The new value for $i$ was determined to be $i=22 \pm 11^\circ$ and the magnetic obliquity angle $\beta$, assuming a dipolar field, was found to be $\beta=77 \pm 11^\circ$.

From our analysis, we deduce that the results are representative of a fossil field. By comparing Stokes V profiles at the same rotational phase, but with observation periods separated by 6 months or more, as is the case in Fig.~\ref{fig:compplot}, we observe very similar features, i.e. the field is stable over a period of at least a year. In addition, HD\,41641 is hot and not a fast rotator, and thus fits the criteria for a typical fossil field-type magnetic field. However, from our modelling, we observed no evidence of a dipolar structure to the field, which points towards a complex fossil field. Such complex fossil field are rare but exists, e.g. HD\,37776 \citep{kochukhov2011}.

Additional observations with better time coverage of this target would allow for greater accuracy in the magnetic field measurements, and the possibility to perform a complex field modelling as was done for HD\,37776. 

\section*{Acknowledgements}
We thank Oleg Kochukhov and Tanya Ryabchikova for useful discussions. This research has made use of the SIMBAD database operated at CDS, Strasbourg (France), and of NASA's Astrophysics Data System (ADS). 
A.E. acknowledges funding from the Fonds voor Wetenschappelijk Onderzoek Vlaanderen (FWO) under contract ZKD1501-00-W01.

\section*{Data Availability}
The data underlying this article are available in the PolarBase database, at \url{https://polarbase.irap.omp.eu}

\pagebreak




\bibliographystyle{mnras}
\bibliography{bibliography} 


\pagebreak
\appendix

\section{Log of observations}

On the following pages is displayed the complete list of Narval observations of HD\,41641, with indications of Julian Date at midpoint of each observation, exposure time and phase assuming F7 as the rotation frequency.

\begin{tabular}[t]{lllll}
\hline
\#  & Date      & Mid-HJD   & $T_{\rm exp}$ & Phase   \\
    &           & -2450000  & s      &         \\
\hline
1   & 21 Oct 15 & 7317.6795 & 4x27   & 0.242 \\
2   &           & 7317.6841 & 4x27   & 0.244 \\
3   &           & 7317.6887 & 4x27   & 0.245 \\
4   &           & 7317.6934 & 4x27   & 0.247 \\
5   &           & 7317.6980 & 4x27   & 0.248 \\
6   &           & 7317.7027 & 4x27   & 0.250 \\
7   &           & 7317.7073 & 4x27   & 0.252 \\
8   &           & 7317.7119 & 4x27   & 0.253 \\
9   & 9 Nov 15  & 7336.7479 & 4x27   & 0.030 \\
10  &           & 7336.7583 & 4x27   & 0.034 \\
11  &           & 7336.7671 & 4x27   & 0.037 \\
12  & 11 Nov 15 & 7338.7031 & 4x27   & 0.726 \\
13  &           & 7338.7200 & 4x27   & 0.732 \\
14  &           & 7338.7348 & 4x27   & 0.738 \\
15  &           & 7338.7474 & 4x27   & 0.742 \\
16  &           & 7338.7576 & 4x27   & 0.746 \\
17  & 12 Nov 15 & 7339.7383 & 4x27   & 0.095 \\
18  &           & 7339.7429 & 4x27   & 0.096 \\
19  &           & 7339.7475 & 4x27   & 0.098 \\
20  &           & 7339.7522 & 4x27   & 0.100 \\
21  &           & 7339.7568 & 4x27   & 0.101 \\
22  &           & 7339.7615 & 4x27   & 0.103 \\
23  &           & 7339.7661 & 4x27   & 0.105 \\
24  &           & 7339.7708 & 4x27   & 0.106 \\
25  & 16 Nov 15 & 7343.6435 & 4x27   & 0.485 \\
26  &           & 7343.6481 & 4x27   & 0.487 \\
27  &           & 7343.6528 & 4x27   & 0.488 \\
28  &           & 7343.6574 & 4x27   & 0.490 \\
29  &           & 7343.6621 & 4x27   & 0.492 \\
30  &           & 7343.6668 & 4x27   & 0.493 \\
31  &           & 7343.6716 & 4x27   & 0.495 \\
32  &           & 7343.6764 & 4x27   & 0.497 \\
33  & 15 Mar 16 & 7463.3214 & 4x27   & 0.090 \\
34  &           & 7463.3263 & 4x27   & 0.092 \\
35  &           & 7463.3313 & 4x27   & 0.094 \\
36  &           & 7463.3360 & 4x27   & 0.096 \\
37  &           & 7463.3410 & 4x27   & 0.097 \\
38  &           & 7463.3459 & 4x27   & 0.099 \\
39  &           & 7463.3505 & 4x27   & 0.101 \\
40  &           & 7463.3556 & 4x27   & 0.103 \\
41  & 20 Mar 16 & 7468.3462 & 4x27   & 0.879 \\
42  &           & 7468.3508 & 4x27   & 0.881 \\
43  &           & 7468.3557 & 4x27   & 0.883 \\
44  &           & 7468.3605 & 4x27   & 0.884 \\
45  &           & 7468.3652 & 4x27   & 0.886 \\
46  &           & 7468.3703 & 4x27   & 0.888 \\
47  &           & 7468.3752 & 4x27   & 0.900 \\
\hline
\end{tabular}

\begin{tabular}[t]{lllll}
\hline
\#  & Date      & Mid-HJD   & $T_{\rm exp}$ & Phase   \\
    &           & -2450000  & s      &         \\
\hline
48  & 8 Oct 16  & 7670.6194 & 4x27   & 0.888 \\
49  &           & 7670.6228 & 4x27   & 0.890 \\
50  &           & 7670.6261 & 4x27   & 0.891 \\
51  &           & 7670.6294 & 4x27   & 0.892 \\
52  &           & 7670.6327 & 4x27   & 0.893 \\
53  &           & 7670.6360 & 4x27   & 0.894 \\
54  &           & 7670.6394 & 4x27   & 0.896 \\
55  &           & 7670.6427 & 4x27   & 0.897 \\
56  &           & 7670.6460 & 4x27   & 0.898 \\
57  &           & 7670.6494 & 4x27   & 0.899 \\
58  & 27 Oct 16 & 7689.5618 & 4x27   & 0.632 \\
59  &           & 7689.5651 & 4x27   & 0.633 \\
60  &           & 7689.5684 & 4x27   & 0.634 \\
61  &           & 7689.5718 & 4x27   & 0.636 \\
62  &           & 7689.5751 & 4x27   & 0.637 \\
63  &           & 7689.5784 & 4x27   & 0.638 \\
64  &           & 7689.5818 & 4x27   & 0.639 \\
65  &           & 7689.5851 & 4x27   & 0.640 \\
66  &           & 7689.5886 & 4x27   & 0.642 \\
67  &           & 7689.5919 & 4x27   & 0.643 \\
68  & 28 Oct 16 & 7690.6884 & 4x27   & 0.033 \\
69  &           & 7690.6918 & 4x27   & 0.034 \\
70  &           & 7690.6952 & 4x27   & 0.035 \\
71  &           & 7690.6985 & 4x27   & 0.037 \\
72  &           & 7690.7018 & 4x27   & 0.038 \\
73  &           & 7690.7052 & 4x27   & 0.039 \\
74  &           & 7690.7085 & 4x27   & 0.040 \\
75  &           & 7690.7121 & 4x27   & 0.042 \\
76  &           & 7690.7158 & 4x27   & 0.043 \\
77  &           & 7690.7192 & 4x27   & 0.044 \\
78  & 29 Oct 16 & 7691.5907 & 4x27   & 0.354 \\
79  &           & 7691.5940 & 4x27   & 0.355 \\
80  &           & 7691.5974 & 4x27   & 0.357 \\
81  &           & 7691.6008 & 4x27   & 0.358 \\
82  &           & 7691.6041 & 4x27   & 0.359 \\
83  &           & 7691.6075 & 4x27   & 0.360 \\
84  &           & 7691.6109 & 4x27   & 0.361 \\
85  &           & 7691.6142 & 4x27   & 0.363 \\
86  &           & 7691.6175 & 4x27   & 0.364 \\
87  &           & 7691.6208 & 4x27   & 0.365 \\
88  & 30 Oct 16 & 7692.7180 & 4x27   & 0.756 \\
89  &           & 7692.7214 & 4x27   & 0.757 \\
90  &           & 7692.7247 & 4x27   & 0.758 \\
91  &           & 7692.7281 & 4x27   & 0.759 \\
92  &           & 7692.7315 & 4x27   & 0.760 \\
93  &           & 7692.7349 & 4x27   & 0.762 \\
94  &           & 7692.7382 & 4x27   & 0.763 \\
95  &           & 7692.7416 & 4x27   & 0.764 \\
96  &           & 7692.7449 & 4x27   & 0.765 \\
97  &           & 7692.7482 & 4x27   & 0.766 \\
\hline
\end{tabular}
\begin{table}
\begin{tabular}[t]{lllll}
\hline
\#  & Date      & Mid-HJD   & $T_{\rm exp}$ & Phase   \\
    &           & -2450000  & s      &         \\
\hline
98  & 31 Oct 16 & 7693.6122 & 4x27   & 0.074 \\
99  &           & 7693.6156 & 4x27   & 0.075 \\
100 &           & 7693.6189 & 4x27   & 0.076 \\
101 &           & 7693.6223 & 4x27   & 0.078 \\
102 &           & 7693.6256 & 4x27   & 0.079 \\
103 &           & 7693.6291 & 4x27   & 0.080 \\
104 &           & 7693.6325 & 4x27   & 0.081 \\
105 &           & 7693.6359 & 4x27   & 0.082 \\
106 &           & 7693.6392 & 4x27   & 0.084 \\
107 &           & 7693.6426 & 4x27   & 0.085 \\
108 & 2 Nov 16  & 7695.6428 & 4x27   & 0.797 \\
109 &           & 7695.6462 & 4x27   & 0.798 \\
110 &           & 7695.6496 & 4x27   & 0.799 \\
111 &           & 7695.6530 & 4x27   & 0.800 \\
112 &           & 7695.6564 & 4x27   & 0.802 \\
113 &           & 7695.6597 & 4x27   & 0.803 \\
114 &           & 7695.6631 & 4x27   & 0.804 \\
115 &           & 7695.6664 & 4x27   & 0.805 \\
116 &           & 7695.6700 & 4x27   & 0.807 \\
117 &           & 7695.6736 & 4x27   & 0.808 \\
118 & 25 Nov 16 & 7718.6157 & 4x27   & 0.975 \\
119 &           & 7718.6190 & 4x27   & 0.976 \\
120 &           & 7718.6223 & 4x27   & 0.978 \\
121 &           & 7718.6256 & 4x27   & 0.979 \\
122 &           & 7718.6289 & 4x27   & 0.980 \\
123 &           & 7718.6323 & 4x27   & 0.981 \\
124 &           & 7718.6356 & 4x27   & 0.982 \\
125 &           & 7718.6389 & 4x27   & 0.983 \\
126 &           & 7718.6422 & 4x27   & 0.985 \\
127 &           & 7718.6455 & 4x27   & 0.986 \\
128 & 1 Dec 16  & 7724.4702 & 4x27   & 0.059 \\
129 &           & 7724.4736 & 4x27   & 0.061 \\
130 &           & 7724.4769 & 4x27   & 0.062 \\
131 &           & 7724.4802 & 4x27   & 0.063 \\
132 &           & 7724.4836 & 4x27   & 0.064 \\
133 &           & 7724.4869 & 4x27   & 0.065 \\
134 &           & 7724.4902 & 4x27   & 0.067 \\
135 &           & 7724.4935 & 4x27   & 0.068 \\
136 &           & 7724.4967 & 4x27   & 0.069 \\
137 &           & 7724.5001 & 4x27   & 0.070 \\
138 & 2 Dec 16  & 7725.6019 & 4x27   & 0.462 \\
139 &           & 7725.6052 & 4x27   & 0.463 \\
140 &           & 7725.6086 & 4x27   & 0.465 \\
141 &           & 7725.6120 & 4x27   & 0.466 \\
142 &           & 7725.6153 & 4x27   & 0.467 \\
143 &           & 7725.6186 & 4x27   & 0.468 \\
144 &           & 7725.6220 & 4x27   & 0.469 \\
145 &           & 7725.6254 & 4x27   & 0.471 \\
146 &           & 7725.6287 & 4x27   & 0.472 \\
147 &           & 7725.6323 & 4x27   & 0.473 \\
\hline
\end{tabular}
\caption{Journal of Narval observations of HD\,41641, with indications of Julian Date at midpoint of each observation, exposure time and phase according to F7.}
\label{tab:observations2}
\end{table}

\bsp	
\label{lastpage}
\end{document}